\documentclass[12pt,a4paper]{article}
\usepackage[dvips]{graphicx}
\usepackage[textwidth=16cm,textheight=23cm]{geometry}
\usepackage{pgf}
\usepackage{hyperref}
\usepackage{amsmath}
\usepackage[square,numbers]{natbib}
\usepackage{listings}
\usepackage[right]{lineno}

\newcommand{\mycode}[1]{\lstinline!#1!}

\begin{document}

\title{\texttt{automan}: a simple, Python-based, automation framework for numerical computing}
\author{Prabhu Ramachandran\footnotemark[1]}

\footnotetext[1]{Department of Aerospace Engineering, Indian Institute of
  Technology Bombay, Powai, Mumbai 400076}

\maketitle


\begin{abstract}
  We present an easy-to-use, Python-based framework that allows a researcher
  to automate their computational simulations. In particular the framework
  facilitates assembling several long-running computations and producing
  various plots from the data produced by these computations. The framework
  makes it possible to reproduce every figure made for a publication with a
  single command. It also allows one to distribute the computations across a
  network of computers. The framework has been used to write research papers
  in numerical computing. This paper discusses the design of the framework,
  and the benefits of using it. The ideas presented are general and should
  help researchers organize their computations for better reproducibility.
\end{abstract}

\section{Introduction}
\label{sec:intro}

It is well known that reproducibility is a cornerstone of science. Clearly,
reproducibility (or repeatability) is vitally important for computational
science as well. \citet{lbarba:reproducibility-syllabus} provides a succinct
overview of some of most important articles on reproducibility. Ensuring
repeatability in computational work takes additional effort. Unfortunately,
there are not many immediate or direct incentives for a researcher to invest
time in building systems that ensure repeatability. Further, as discussed in
\cite{repro:cfd:mesnard:cise:2017}, there can be significant challenges
involved in carefully repeating and replicating numerical experiments.

In this paper we discuss how we automated a computational research paper. We
believe that this is a very important step in facilitating reproducible
research. The approach used in our automation framework is fairly general and
could be of use to other researchers. Our implementation is open source and
available at \url{http://github.com/pypr/automan}. What is perhaps more
important is that we discovered that repeatability can be very profitable to
the researcher. This is important since it provides an incentive for one to
invest time in making computational research repeatable. It is our hope that
the ideas discussed in the current work and our framework are of general use
and facilitate greater reproducibility in computational science.

Many computational science papers tend to involve the following tasks:
\begin{itemize}

\item run several computations, each requiring several hours to complete. The
  programs executed are typically compiled programs written in a low-level
  language like FORTRAN, C, or C++. Sometimes these may be implemented in a
  higher-level language like Python or MATLAB. Regardless of the choice of
  programming language, researchers often need to run different programs with
  various parameters.

\item compare results generated by different computational methods and compare
  computational results with theoretical or experimental data.

\item collate these results into a variety of plots, tables, and other forms
  suitable for sharing in the form of a publication.

\end{itemize}
Once these plots and tables are generated, the authors can complete a
manuscript describing the novelty of the study along with the results of the
study.

Our automation framework, called \texttt{automan}, was used to automate the
computations performed during the development of a Smoothed Particle
Hydrodynamics (SPH) scheme that we implemented called the ``Entropically
Damped Artificial Compressibility SPH'' \cite{PRKP:EDAC-SPH:2016}, the scheme
is henceforth called EDAC-SPH.

SPH is a particle-based scheme that can be used to simulate a wide variety of
problems (see \cite{monaghan-review:2005} for a review). There are many SPH
``schemes'' in the literature. In our work, we needed to compare our results
with established SPH schemes as well as known exact solutions and experimental
data. The framework allowed us to:
\begin{itemize}
\item run all simulations and produce all the figures for the manuscript with
  a single command.
\item incrementally add, modify, and rerun simulations.
\item easily plot and compare a variety of similar simulations.
\item modify and update any of the figures and plots without needing to re-run
  the long-running simulations.
\item distribute our simulations on a collection of other idle desktop
  computers on the network.
\end{itemize}

The framework is implemented entirely in Python~\cite{py:python}. There are
several automation frameworks that already exist. We had initially used
pydoit~\cite{doit} and later luigi~\cite{luigi}. Both of these tools are
certainly very useful and many of the ideas used in the current framework are
inspired from there. Lancet~\cite{lancet:2013} provides very interesting
abstractions of how to parametrize, and specify simulations, however, it does
not offer any task management or automation facilities.

The venerable ReDoc~\cite{redoc:schwab:cise:2002} offers a general strategy
to automate tasks using a Makefile with a small amount of custom rules that
they provide. These allow a scientist to create, view, delete any figures
produced and also clean any intermediate results produced. They also suggest
that researchers segregate their files into three different categories of
files based on how easy they are to recreate. The ideas provided by ReDoc are
very important, however, the details of the implementation are not always
optimal for long running computations on loosely distributed computers. Make
is best suited for compiling programs and the execution and scheduling of the
tasks is not easy to control or modify. Further, ReDoc does not help abstract
key tasks that are important when analyzing computational simulations. Our
framework attempts to address some of these.

Sumatra~\cite{sumatra-davison-2014}, takes a more comprehensive approach by
providing a suite of commands to capture the details of all computational
experiments including the versions of the dependent packages that are used to
run every command. This provides a much more comprehensive framework. Our
framework is orthogonal to the goals of Sumatra and makes it easier to manage
a collection of tasks. We do not capture the entire environment or track the
results themselves. As such, it is possible to use the sumatra framework with
ours.

Nextflow~\cite{nextflow-nature-biotech-2017} is a very powerful package for
reproducibility. It provides a domain specific language for specifying, and
executing computational pipelines. It can optionally use docker containers to
provide repeatability in computational simulations. It interfaces well with
existing pipelines written in other scripting languages. Our approach is much
simpler, and completely implemented in Python. Given that our analysis and
post-processing code is written in Python, our framework is a more natural fit
for us. In addition our framework provides some general abstractions to group
and split the simulations and avoid repetitive code.

Our framework provides the following core features:
\begin{itemize}
\item the ability to define arbitrary tasks and their dependencies in Python.
  This is exactly similar to what is provided by the luigi
  package~\cite{luigi} mentioned above.
\item the ability to perform post-processing using the convenience of Python.
  The framework provides several convenient abstractions that make this easier
  to do.
\item minimize repetitive code for comparison of similar computational
  schemes.
\item control the scheduling and execution of tasks, including running tasks
  on remote computers.
\end{itemize}

We point out several common patterns and abstractions that were useful in
implementing our framework. These ideas are general and could allow other
researchers to better organize their own computations for reproducibility.

In the following, we use examples from our computations in order to
demonstrate the automation framework. The remainder of the paper is organized
as follows. We provide a brief, high-level background of our research paper in
order to provide a context for the automation framework. We then discuss the
overall design of our framework. We discuss how the framework allowed us to
distribute the computations across generic Linux/MacOS computers on the
network accessible via \verb+ssh+. We discuss how our implementation provided
features that were not readily available in many other tools and discuss
possible future directions of the framework.

\section{Background}
\label{sec:background}

As mentioned in the introduction, computational work often requires executing
programs that run for a long period of time. This can be anywhere from a few
hours to several days and sometimes weeks. The programs are often to be run
with different initial conditions or parameters in order to explore a variety
of questions. After these simulations are run, researchers need to compare the
results of these computations and produce plots and tables that are assembled
into a manuscript.

Typically these individual simulations are run manually, thereafter, plots and
tables are made manually and then added to a manuscript. This workflow does
not scale and does not lend itself for easy reproducibility. The
recommendations of Wilson et
al.~\cite{best-practices-sci-comp:wilson:plos1:2014} and
\citet{repro:10rules:sandve:pcbi:2013} provide very general guidelines to make
this workflow more productive, reliable, and reproducible.

In this paper we use the Smoothed Particle Hydrodynamics method as a typical
example of a numerical method. As with any established numerical method, there
are different variants of the basic method. In the following we use the term
``scheme'' to denote a variant of a basic numerical method. In our research
work~\cite{PRKP:EDAC-SPH:2016}, a new SPH scheme called EDAC-SPH was being
developed to simulate incompressible fluid flow problems. This scheme was
compared with an established scheme called the Weakly-Compressible SPH
formulation (WCSPH) as well as a recent SPH formulation called the
Transport-Velocity Formulation (TVF).

The entire implementation of the WCSPH, TVF, and new schemes was made using
the PySPH framework~\cite{PR:pysph:scipy16,pysph}. The PySPH framework allows
users to write the SPH schemes and programs in the Python programming
language. Each program solving a benchmark problem, written using PySPH,
supports a variety of command line arguments to configure various parameters.
Users are encouraged to add their own command line options to their scripts.
While all of the examples in this work use Python, it is important to note
that one may use any executable program or script with automan.

A simple workflow for performing the necessary numerical simulations would
involve the following steps:
\begin{enumerate}
\item Break-up each benchmark problem into several command line program
  invocations that can be executed independently. Each of these simulations is
  typically called a ``case''.
\item For each benchmark problem, execute each of the simulations (cases).
\item Once the simulations are completed, gather the post-processed data or
  perform additional post-processing and create any plots or tables.
\item Assemble the relevant plots for the manuscript.
\end{enumerate}

The above workflow works best when the programs support command line arguments
to setup various parameters. This can be extended to cases which require
specific input files as well. The framework requires that each case produce
any output files in a specified and configurable directory.

We consider a few example problems to clarify the above. In the case of our
research, one of the problems we simulated is the Taylor-Green-Vortex problem
(TGV) which has a known exact solution. We needed to simulate several cases of
this problem with nine different variants of SPH schemes. These results needed
to be compared. The command line programs for these would typically be of the
following form:
\begin{itemize}

\item Use the WCSPH scheme as implemented in the pysph package (note that
  the command ``\mycode{pysph run}'' executes a standard PySPH example):
\begin{verbatim}
$ pysph run taylor_green --scheme wcsph --nx 50 -d wsph
\end{verbatim}

  \item Use the WCSPH scheme and add a tensile correction with the default
    value of \mycode{nx}:
\begin{verbatim}
$ pysph run taylor_green --scheme wcsph --tensile-correction -d tc
\end{verbatim}

  \item Use the TVF scheme (the code is a Python script in this case):
\begin{verbatim}
$ python taylor_green.py --scheme tvf --nx 50 -d tvf
\end{verbatim}

\end{itemize}
The ``\mycode{-d path}'' argument ensures that the results are written to the
specified path. Once each of these commands complete, the results of the
simulation along with any post-processing are made available inside the
respective directories. Typically, PySPH scripts create a \mycode{results.npz}
file which contains any post-processed results in raw form. The \mycode{.npz}
extension is used to store standard NumPy arrays. These results then need to
be compared and plotted. Sometimes new post-processing is necessary for the
purposes of a manuscript that may not be relevant to the case itself. This
represents part of a single benchmark problem being explored. There will
typically be many more cases and benchmark problems to simulate.

Clearly, command line arguments are used extensively here. If the user has to
create different input files to run these simulations, that could also be
supported by the framework. However, having extensive command line argument
support greatly facilitates the automation of the simulations.

In the next section we describe how the new framework can be used to automate
the execution and post-processing of these simulations.

\section{Design of the automation framework}
\label{sec:design}

\begin{figure}[ht]
  \centering
  \includegraphics[width=0.8\textwidth]{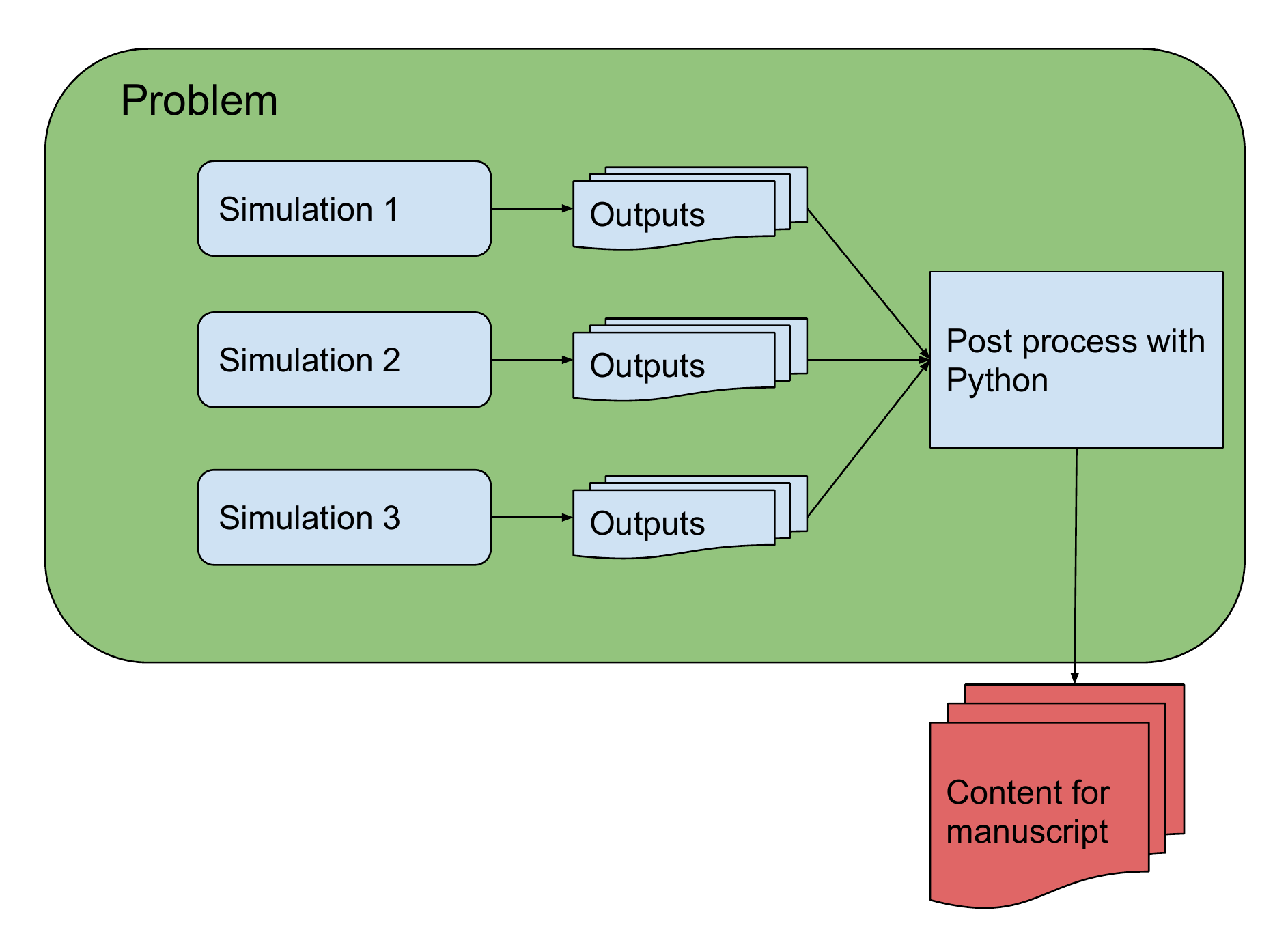}
  \caption{Sketch illustrating how a Problem encapsulates several simulations
    and some post-processing which produces content for the manuscript.}
  \label{fig:problem_design}
\end{figure}

\begin{figure}[h!]
  \centering
  \includegraphics[width=0.8\textwidth]{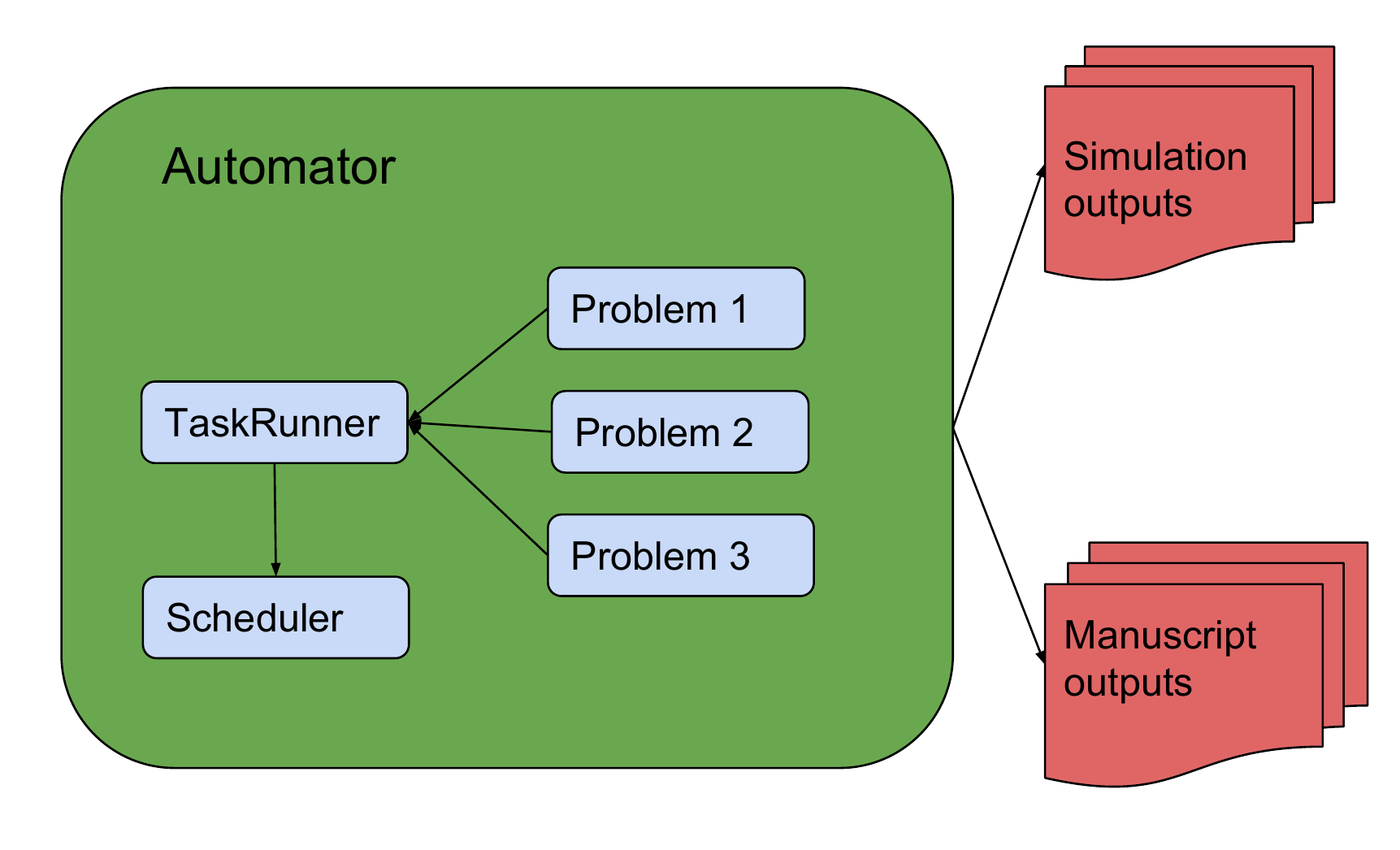}
  \caption{Sketch of the overall design of the automator which manages several
    problems and executes the simulations.}
  \label{fig:automator_design}
\end{figure}

The general approach to using the framework is as follows:
\begin{itemize}
\item Each set of related simulations are grouped into what is called a
  ``Problem''. Each problem also contains Python code for producing different
  plots from these simulations. Users create subclasses of a base ``Problem''
  class to customize the required simulations and analysis for that particular
  problem. This is illustrated in Fig~\ref{fig:problem_design}. The
  post-processing generates content that can be immediately used in a
  manuscript.
\item Each simulation to be run can be created as either an instance of a
  ``Simulation'' class or specified as a command string to be executed. The
  simulation class makes it very easy to abstract common plotting tasks for a
  given kind of simulation.
\item All the problem classes are collected and passed to an ``Automator''
  instance which is given two directory paths, one for the simulation outputs
  and one for the final manuscript figures. This object checks if the final
  outputs are already made, if not, it runs any necessary simulations (if
  their output is not already created) and the Python code for producing the
  plots. The simulations can be run either on the local machine or distributed
  across a cluster of machines. This is done by delegating them to a task
  runner which in turn delegates the actual execution to a scheduler. This is
  illustrated in Fig~\ref{fig:automator_design}.
\end{itemize}

\subsection{Simple example}
\label{sec:simple-example}

We consider a simple example to elaborate the design discussed in
section~\ref{sec:design}. Listing~\ref{lst:simple-example} shows the outline
of the contents of a file, \mycode{automate.py}, illustrating the use of the
automation framework.
\lstset{numbers=left}
\begin{lstlisting}[label=lst:simple-example,caption={\mycode{automate.py}
illustrating the automation framework.}]
from automan.api import Problem, Simulation, Automator

class TaylorGreen(Problem):
    def get_name(self):
        return 'taylor_green'

    def setup(self):
        pysph_cmd = 'pysph run taylor_green'
        self.cases = [
            Simulation(
                root=self.input_path('std_sph'),
                base_command=pysph_cmd,
                job_info=dict(n_core=1, n_thread=2),
                scheme='wcsph', nx=50
            ),
            Simulation(
                root=self.input_path('std_sph_tensile_correction'),
                base_command=pysph_cmd,
                job_info=dict(n_core=1, n_thread=2),
                scheme='wcsph', nx=50, tensile_correction=None
            ),
            # ...
        ]

    def run(self):
        self.make_output_directory()
        self._plot_decay_error_all()
        self._plot_decay_error()

automator = Automator(
    simulation_dir='outputs',
    output_dir='manuscript/figures',
    all_problems=[TaylorGreen]
)
automator.run()
\end{lstlisting}

We start by looking at the \mycode{automator} object (lines 30--34). It is
passed two directory names, \mycode{'outputs'} and
\mycode{'manuscript/figures'} as well as a list of Problem classes. The
\mycode{outputs} directory will contain the outputs generated by the
simulations and the other will contain the final plots for the manuscript. The
\mycode{automator} instance has a \mycode{run} method that is executed to
perform the automation. Several things can be noted in the
\mycode{TaylorGreen} class (starting at line 3) which is a subclass of the
\mycode{Problem} class.
\begin{itemize}
\item The \mycode{get_name} method (lines 4-5) returns a name which is a
  sub-directory containing all relevant outputs for this particular problem.
  When this problem is executed, all the simulation outputs will be inside
  \mycode{outputs/taylor_green/}. The figures will be inside
  \mycode{manuscript/figures/taylor_green/}.
\item The \mycode{setup} method (lines 7-23) simply creates a collection of
  \mycode{cases}. Each case is an instance of a \mycode{Simulation} class.
  Each simulation is given a root path and a basic command along with some
  information on the number of cores and threads desired. The \mycode{n_core}
  argument is used to distribute the load to different machines depending on
  the availability of free cores. The number of threads is a parameter used by
  OpenMP. All subsequent keyword arguments are converted to command line
  arguments automatically.

  The simulation instances are passed additional options explicitly as keyword
  arguments in Python instead of in the command string to make it easy to
  filter cases for post-processing. For example, it is possible to easily
  filter all the cases which have \mycode{nx=50} and \mycode{scheme='wcsph'}.
  This is discussed later.

\item The \mycode{run} method (lines 25--28) performs any post-processing of
  the outputs in order to produce the final figures for the manuscript. The
  details of the methods \mycode{self._plot_decay_error_all()} etc.\ (lines
  27, 28) are not discussed here. They are simple functions that make any
  necessary plots. The simulation instances make it easy to refer to files
  within the simulation output.
\end{itemize}

The \mycode{automate.py} completely specifies the simulations. The plotting
code is routine Python (and not shown in the listing). To use the automation
script, a user runs:
\begin{verbatim}
    $ python automate.py
\end{verbatim}
The automator first checks if \mycode{manuscript/figures/taylor_green}
directory exists. If it does not, all the simulations are executed and the
plotting code is called automatically after the simulations complete.

It is often necessary to modify or change the plots without re-running the
simulations, this can be achieved by executing:
\begin{verbatim}
    $ python automate.py -f
\end{verbatim}
To re-run completed simulations, one must explicitly delete the appropriate
directories in the \mycode{outputs} directory. On the other hand, if the
simulation code itself fails, one may simply correct the simulation code and
re-run the automation script, and the corrected code will be executed again.

In similar fashion, a user could create any number of \mycode{Problem} classes
and all of their simulations and post-processing will be executed if these
classes are passed to the automator (see line 33 in
Listing~\ref{lst:simple-example}).

When the \mycode{automate.py} script is executed, it simulates all problems by
default. This is sometimes inconvenient when one wishes to only create the
plots for a particular problem, this can be done by specifying the particular
problem, for example:
\begin{verbatim}
    $ python automate.py -f TaylorGreen EllipticalDrop
\end{verbatim}
will execute only the Taylor-Green and elliptical drop problems. The name of
the problem specified is not case sensitive.

When working actively on research, there are times when one may wish to run
just one particular simulation out of the many for a problem.  In this case,
one can do the following:
\begin{verbatim}
    $ python automate.py TaylorGreen -m "*tensile*"
\end{verbatim}

\begin{sloppypar}
  This will run just a specific set of simulations that match the criterion,
  and in this case it will run the \mycode{std_sph_tensile_correction} case.
  It will not execute the post-processing code since the other cases may not
  have been run. This is particularly useful if one only wants to run a few
  simulations without generating all the plots for the problem. The search
  criterion is matched using standard Unix \mycode{fnmatch} patterns to make
  it easy to select specific cases to run.
\end{sloppypar}

\subsection{More details on simulations and problems}

The \mycode{Simulation} and \mycode{Problem} classes are clearly important in
the above example and these classes provide some convenient functionality.
Creating a \mycode{Simulation} object is generally done as follows:
\lstset{numbers=none}
\begin{lstlisting}
>>> s = Simulation(root_dir, base_command, job_info, **kw)
\end{lstlisting}

The simulation object stores a path (\mycode{root_dir}) to where the
simulation output should be generated and is given a basic command line to
run. The argument \mycode{job_info} is optional and helps specify the
execution context in terms of number of cores and threads needed by the
simulation. Any additional keyword arguments are converted into command line
arguments using the \mycode{get_command_line_args} method of the class. This
makes it easy to customize the command line arguments. There are other methods
provided to facilitate convenient access to the data generated by the
simulation and making plots from the outputs. These are illustrated by way of
an example below (on an interactive Python interpreter):
\lstset{numbers=none}
\begin{lstlisting}[label=lst:sim-py,caption={Illustration of
\mycode{Simulation} class usage and behavior.}]
>>> s = Simulation(
...     'outputs/sph',
...     'pysph run elliptical_drop',
...     job_info=dict(n_core=1, n_thread=2),
...     timestep=0.005,
...     tensile_correction=None
... )

>>> s.name
'sph'

>>> s.command
'pysph run elliptical_drop --timestep=0.005 --tensile-correction'

>>> s.get_command_line_args()
'--timestep=0.005 --tensile-correction'

>>> s.input_path('results.npz')
'outputs/sph/results.npz'

>>> s.data.fid
<open file 'outputs/sph/results.npz' ...>

>>> s.render_parameter('tensile_correction')
'tensile_correction'

>>> s.get_labels(['timestep'])
'timestep=0.005'
\end{lstlisting}

Note that each simulation has a \mycode{name} which is the name of directory
in which the output is generated. The \mycode{command} property has
automatically converted the parameters \mycode{timestep, tensile_correction}
to suitable command line arguments. It is important to note again that the
commands that are executed such that one may configure the output directory in
which they will generate output using a command line argument. In the present
case, this additional command line argument is added by the \mycode{PySPHTask}
discussed later below. If a user wishes to specify this in the simulation one
could use the special string \mycode{\$output_dir} to specify the directory
and this will be automatically substituted. For example one could specify the
simulation as follows: \lstset{numbers=none}
\begin{lstlisting}
>>> s = Simulation('outputs/sph', 'command -d $output_dir')
\end{lstlisting}
Here, the \mycode{\$output_dir} will be substituted with the actual directory
name when the simulation is executed by a task.

The \mycode{input_path} method makes it easy to refer to a path inside the
output directory. The \mycode{render_parameter} method returns a string for
each parameter given to it. This is convenient when one wishes to ensure that
when a legend is rendered for a parameter (say \mycode{alpha} which is to be
rendered as $\alpha$) that it uses a suitable LaTeX label. A user may choose
to configure it to their needs. The \mycode{get_labels}, renders out multiple
such parameters suitable for a legend.

The class also provides a \mycode{data} property that returns the data loaded
from a \mycode{results.npz} file generated by many PySPH examples. Clearly,
this will need to be configured to suit other packages. This property makes it
easy to plot post-processed results.

When many plots needed to be made from a given kind of simulation, it is
convenient to create a customized subclass of the \mycode{Simulation} class.
For the paper in \cite{PRKP:EDAC-SPH:2016}, there were around 45 Taylor-Green
simulations to make. This required a custom subclass as below:
\lstset{numbers=left}
\begin{lstlisting}[label=lst:sim-extend, caption={Example illustrating a custom
subclass of the \mycode{Simulation}}.]
class TGV(Simulation):
    def get_command_line_args(self):
        # ...
    def render_parameter(self, param):
        # ...
    def l1(self, **kw):
        import matplotlib.pyplot as plt
        data = self.data
        plt.plot(data['t'], data['l1'], **kw)
        plt.xlabel('t'); plt.ylabel(r'$L_1$ error')
\end{lstlisting}
This implementation overrides the default to customize how the command line
arguments are generated. The parameter rendering is also changed from the
default. Several convenient plotting methods are implemented and we show only
one \mycode{l1} (line 6 above). This method plots the $L_1$ error using
\mycode{matplotlib}.

Since the parameters of interest are passed to a simulation as additional
keyword arguments, it is possible to filter out specific cases for comparison.
The automation framework provides a few convenient functions. For example,
take the case where we wish to plot the $L_1$ error for cases with different
numerical schemes but where \mycode{nx=50}. Recall that the actual plots are
made by the \mycode{Problem} class, in our example, this is the
\mycode{TaylorGreen} class in listing~\ref{lst:simple-example} and the code
for the plots would be:
\begin{lstlisting}[label=lst:filter-example, caption={Example illustrating
convenient functions for post-processing simulations.}]
    def _plot_decay_error_all(self):
        import matplotlib.pyplot as plt

        cases = filter_cases(self.cases, nx=50, perturb=0.2)

        plt.figure()
        compare_runs(cases, 'l1', labels=['scheme'])

        plt.legend(loc='upper left')
        plt.savefig(self.output_path('l1_error_all.pdf'))
        plt.close()
\end{lstlisting}

There are two new functions introduced here:
\begin{itemize}
\item \mycode{filter_cases}: this function is given a sequence of cases and
  any additional parameters (see line 4 above). It filters out the cases that
  satisfy the particular parameters. In the code above, all the cases with
  \mycode{nx=50} with \mycode{perturb=0.2} are returned.
\item \mycode{compare_runs}: calls the given method name (\mycode{'l1'} in
  line 7 above) for all the cases and labels them with the given set of labels
  to be indicated in the plot. It can also be passed a method name for an
  exact solution plot. Instead of method names, any callable function can be
  passed. When this function is called, it is passed the simulation instance.
\end{itemize}
An additional convenience function is \mycode{filter_by_name} which picks the
cases given particular names.

These methods make it easy to manage many simulation instances and quickly
produce a publication-ready plot comparing various aspects of the simulations.
In our case, we had one \mycode{Problem} class with 16 instances of the
\mycode{TGV} simulation object and another with 29. The plotting code is not
repeated as it is abstracted out in the \mycode{TGV} class. The filtering
functions along with the \mycode{compare_runs} function allows us to compare
any number of specific simulations very easily. This greatly reduces the
repetitive plotting code for a large number of plots.

The \mycode{Problem} class has been used in listing~\ref{lst:simple-example}
and collects various \mycode{Simulation} instances and specifies how the
simulations are compared to produce a variety of outputs. The problem class has
the following important features:

\begin{itemize}
\item A problem class is always passed the simulation directory and the output
  directory. In our example above, this would correspond to \mycode{'outputs'}
  and \mycode{'manuscript/figures'} respectively. Instantiating the class does
  not result in any computations being executed.

\item The \mycode{get_name} method is used to specify the directory for this
  particular problem which collect both the simulation outputs and the figures.

\item The \mycode{setup} method is used to define the various simulation
  cases. We have already seen this method used in the initial example,
  Listing~\ref{lst:simple-example}.

\item The \mycode{run} method is used to compare the simulations and produce
  the necessary output.  This has also been seen earlier.

\item The \mycode{input_path} and \mycode{output_path} methods are convenience
  methods to help find files in the input and output directories.

\item The \mycode{get_commands} method returns a list of commands to execute
  in order to run the simulations.

\item The \mycode{get_requires} method returns a list of tasks to execute in
  order to run the simulations. Tasks are discussed in the next section and
  allow the user to define the commands to be executed in a powerful way.  By
  default, the tasks are automatically created from the commands.  Users may
  overload this method to create tasks in other ways.

\end{itemize}

The \mycode{Simulation} and \mycode{Problem} classes thus allow the user to
specify the simulations to be executed and how these are to be compared. This
is sufficient information to start using the framework. The next sections
discuss the execution of the tasks as well as the simple cluster management
that is available.

\subsection{Scheduling and execution of tasks}
\label{sec:tasks}

The automation framework provides a simple design for tasks inspired by the
elgant design of Luigi's~\cite{luigi} tasks. A task has three important
methods:
\begin{itemize}
\item \mycode{complete}: this indicates if the task has successfully
  completed. This method should raise an exception if there is an error while
  running the task.
\item \mycode{run}: this executes the task.
\item \mycode{requires}: this produces a sequence of other tasks that should
  be completed before this task may be \mycode{run}.
\end{itemize}
This design allows one to create complex pipelines with dynamic task
generation and dependencies. A \mycode{PySPHTask} class subclasses the
\mycode{Task} class and provides methods that make it easy to execute a PySPH
simulation, and check if a simulation has completed. It also adds a suitable
command line argument to the command in order to have PySPH generate its
output in the correct directory.  A more general \mycode{CommandTask} is
available which can be used by users who are not using PySPH.

\begin{figure}[ht]
  \centering
  \includegraphics[width=0.8\textwidth]{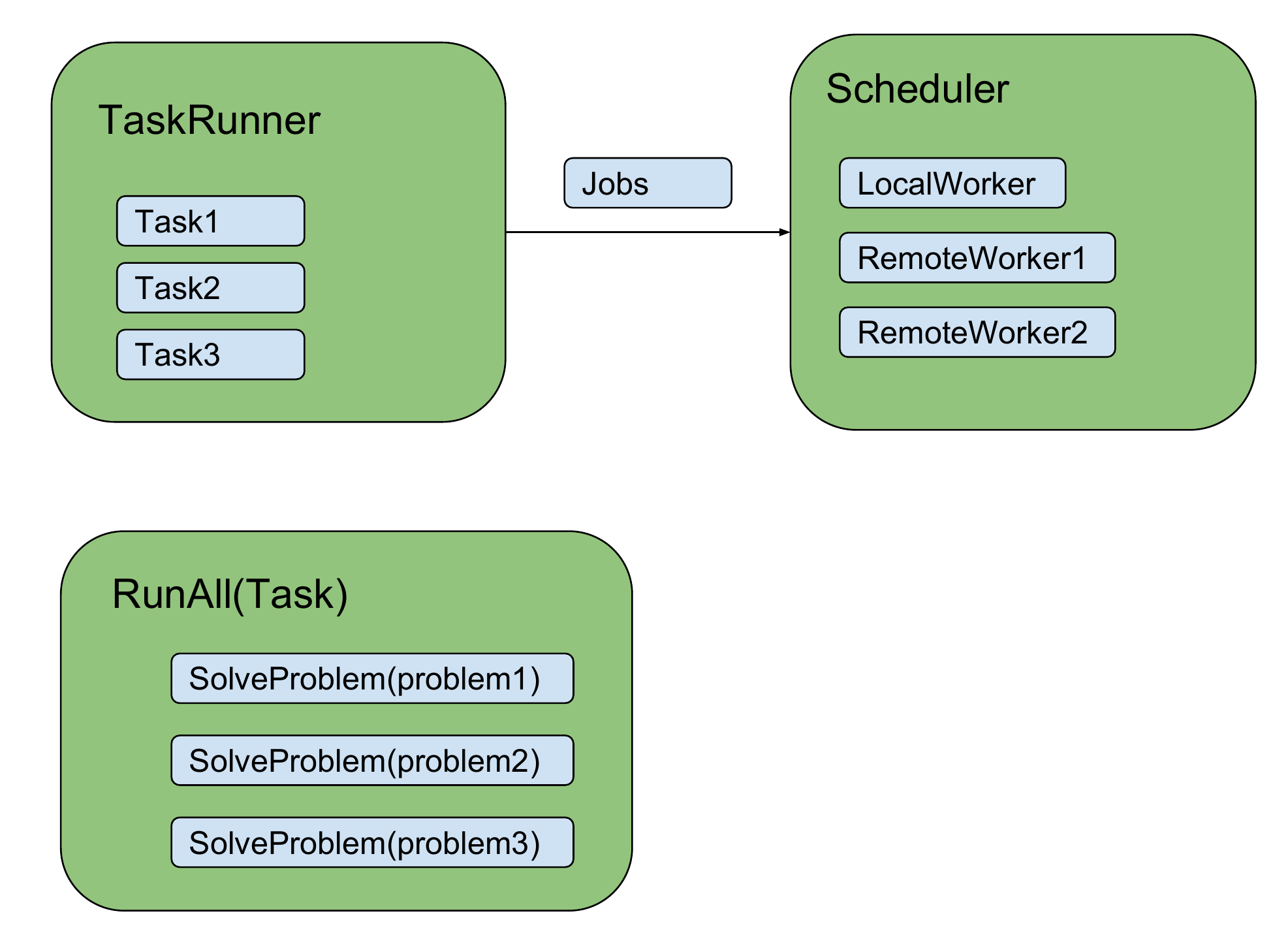}
  \caption{Sketch illustrating the \mycode{TaskRunner} and the
    \mycode{Scheduler}.  Also shown is the \mycode{RunAll} task which manages
    all the given \mycode{Problem} instances.}
  \label{fig:tasks_scheduler}
\end{figure}

A simple \mycode{WrapperTask} is provided which is complete when all of its
required tasks are complete. A \mycode{SolveProblem} task subclass generates
the tasks required to simulate a particular problem and runs the problem's
\mycode{run} method when all the simulations are complete. A \mycode{RunAll}
wrapper task instantiates and executes the given problem classes using the
\mycode{SolveProblem} instances. This is illustrated in
Fig~\ref{fig:tasks_scheduler}.

A \mycode{TaskRunner} class executes the various tasks in the correct order
such that the required tasks are completed first. The tasks are not directly
executed by the \mycode{TaskRunner}. Instead, the execution is passed on to a
\mycode{Scheduler}. A \mycode{Job} class encapsulates the command that needs
to be executed, this includes the environment variables, the number of cores
to use, the output directory etc. A \mycode{Worker} class encapsulates a
computer that may execute the job. A \mycode{LocalWorker} handles local
executions and a \mycode{RemoteWorker} handles remote executions via
\verb+ssh+. A \mycode{Scheduler} manages these workers and is configured using
a simple configuration file. Each time a PySPH task executes, it creates a
\mycode{Job} instance and passes it to the scheduler. The scheduler checks for
any free workers (by checking the CPU load of the computer) and submits the
job to the worker. While tasks are pending or running, \mycode{TaskRunner}
polls the current tasks every so often to see if they are completed. As soon
as tasks are completed, any remaining tasks are scheduled for further
execution. The \mycode{TaskRunner} and \mycode{Scheduler} are illustrated
crudely in the Fig~\ref{fig:tasks_scheduler}.

In this fashion, the tasks are separated from the actual execution of the
programs. This allows one to run the jobs locally or remotely. The outputs
generated on remote computers is copied to the computer where the automate
script is executed. Currently, only local execution and execution via SSH is
supported but it is relatively easy to extend this to support other kinds of
execution contexts.

It is important to note that the expectation from the user is that they have
an executable or script that is configurable via command line arguments and
furthermore that we can specify where it should generate its output files. We
believe that this is not too much to ask from a user. By doing this, we are
able to schedule and execute a large number of simulations and automatically
perform any analysis on the results.

\subsection{Cluster management}

As discussed, the automation framework can transparently execute simulations
on remote machines. The remote machines should be accessible via password-less
SSH. This is easy to do and requires the user to setup an SSH key-pair. The
remote machine should also have the minimum requirements for running PySPH (or
whatever software package one is interested in). For PySPH this requires a C++
compiler and a basic Python installation on the remote machine. Given these,
the user may easily add new remote nodes by running:
\begin{verbatim}
    $ python automate.py -a remote_host
\end{verbatim}
This will copy the necessary files from the local computer to the remote
computer and setup everything on the remote machine. The configuration is
saved to a \mycode{config.json} file. The scripts for setting up and updating
the remote host will be placed in a \mycode{.automan} directory and can be
edited if the remote machines require a different setup.

Once a new computer is added, when one runs the \mycode{automate.py} script it will
automatically distribute the computations across all the workers and the local
machine. If for some reason one does not wish to use a particular worker, one
can remove the entry from the \mycode{config.json} file.

If the source code for the simulations is edited, one can update it on all the
nodes by running:
\begin{verbatim}
    $ python automate.py -u
\end{verbatim}
This updates the sources on all workers, rebuilds PySPH, and runs any problems
specified.

\section{Discussion}

The framework discussed above is fairly simple. The example provided in
section~\ref{sec:simple-example} in Listing~\ref{lst:simple-example} shows
that a user can quickly put together a problem which simulates a variety of
cases and perform some post-processing on the results. The automation
framework ensures the execution of the code in the correct order. Since the
script is written in Python, users may perform any kind of post-processing.
The approach of specifying parameters as keyword arguments allows users to
filter the cases based on the choice of parameters. This facilitates easy
comparison of numerical schemes. By separating the cases being run and the
plotting code, we allow users to compare the cases very easily with little or
no repetition of code. These features are unique to our automation framework.

The Task based infrastructure makes it possible to extend the framework to
other situations with complex dependencies. The task executor
(\mycode{TaskRunner}) and scheduler may also be customized if needed. Most
users are likely to not need this.

As discussed in the introduction, there are many other tools that could have
also been used to perform the dependency and task management. While the other
tools provided inspiration for our implementation, they do not explicitly
provide the additional conveniences that our framework provides. We consider a
few of these tools in the following and discuss why we preferred to implement
our own.

\begin{itemize}
\item Luigi~\cite{luigi} is designed to specify and execute tasks that have
  complex dependencies. It is used to process large amounts of data and is
  very general purpose. While our initial attempts used luigi and some of our
  own design is directly inspired by luigi, we found that it was much easier
  to write our own implementation than use luigi. In particular,
  \begin{itemize}
  \item the default luigi scheduler made it difficult to run multiple jobs in
    parallel in a distributed fashion.
  \item the command line arguments had many additional options that we found
      were confusing to the average user.
  \end{itemize}

\item Doit~\cite{doit} is a very general task management and automation tool.
  However, we found that its scheduler did not always make it easy to
  hierarchically build tasks on demand. On the other hand luigi allowed for
  such an approach. Moreover, it was not clear how a user could to change the
  scheduler easily.

\item Lancet~\cite{lancet:2013} provides general abstractions to decompose
  running a suite of programs into a parametrization of options, specification
  of the actual commands to run, and how to actually execute the task.
  Unfortunately, this does not include any task management or automation of
  tasks and their dependencies.

\end{itemize}

Our framework takes some of the best ideas in these tools to do what we
require. We wish to emphasize that our intention is not to claim that our
solution is universal or necessarily unique. On the other hand, the ideas
discussed are general, related to many other tools, and relatively easy to
implement in a high-level programming language like Python.

ReDoc\cite{redoc:schwab:cise:2002} uses GNU Make which is more well suited to
managing compilation of sources rather than long-time execution and
distribution of tasks. Sumatra~\cite{sumatra-davison-2014} as discussed in the
introduction, is a more general and comprehensive framework.
Nextflow~\cite{nextflow-nature-biotech-2017} is also very comprehensive and
offers many excellent advantages including support for execution via docker
and extensive support for other tools. Nextflow could have perhaps been used
to manage our tasks, however our approach to break up the simulations and
post-processing into Simulation instances and Problem instances is still
useful and could be used with either of these tools.

It is important to note that the \mycode{automan} framework suggests the
following key recommendations,
\begin{itemize}
\item the programs that need to be run should be configurable using command
  line arguments (it is always possible to do this or write wrapper scripts
  for existing programs that do this);
\item the program should be able to generate output into a specified and
  configurable directory.
\item if the programs also perform part of the post-processing, it is often a
  good idea for the post-processing data to be saved into an easy to load
  format for comparison with other executions of the program.
\end{itemize}

As mentioned earlier, spending time on automating our results was very
beneficial. In \cite{PRKP:EDAC-SPH:2016} a new SPH scheme was developed and in
order to test the performance of the new scheme, 11 benchmark problems were
simulated~\footnote{The code and manuscript for the publication is available
  at \url{https://gitlab.com/prabhu/edac_sph/}.}. The results of these were
compared with exact solutions (where available), the results produced by the
WCSPH and TVF implementations, and sometimes compared with results produced by
other numerical methods. This resulted in about 75 different simulations that
required more than 7 days of computational time on a recent quad-core desktop
machine with an Intel i7-4770 CPU at 3.40GHz. The work progressed over the
course of a year while the author had the usual academic workload with a large
amount of time fragmentation. As the work progressed, issues were discovered
that were fixed and several simulations needed to be re-run. Some issues were
major and some were minor but all of these required some re-running of these
simulations. The manuscript was also written incrementally. Once submitted the
reviewer responses were also addressed, sometimes requiring further generation
of plots and additional simulations. Our framework made it very easy to handle
this despite the typical fragmentation of time in academia.

\mycode{automan} is still a very young package and there are some additional
features that are not yet supported. For example, it does not yet support
executing simulations using a batch processing system like torque or slurm. It
has also not been tested with other computational tools apart from PySPH. Our
implementation is open source. Depending on user interest, these features may
be added in the future.

\section{Conclusion}

In this paper we present a simple automation framework called \mycode{automan}
that we use to manage multiple long-running simulations and generate results
suitable for publication. The framework is open source and makes it possible
to run all simulations and generate the figures for publication with one
command. The simulations can be optionally distributed to other computers on a
shared network. An important contribution is a simple abstraction of
separating the different computations into individual problems. Each problem
may require multiple simulations called cases. By explicitly specifying the
parameters of a simulation, we are able to filter and select different cases.
This minimizes repetitive code. The framework helps make a researcher more
productive and facilitates reproducibility.

\section*{Acknowledgments}

We are grateful to Prof.~M.~Ramakrishna for providing valuable feedback on
this manuscript.

\bibliography{references}
\end{document}